\documentstyle[12pt]{article}
\begin{document}
\vspace*{-1.5cm}
{\obeylines
November 1995\hfill PAR--LPTHE 95-11\\[-5mm]
\hfill ULB--TH 05/95 \\[-5mm]
\flushright hep-th/9503202
}
\vskip 1.5cm
\centerline{\bf The Black Hole Entropy Can Be Smaller than {\it
A/}4}
\vskip1cm
\begin{center}
Fran\c{c}ois~Englert
\footnote{ Permanent address: Service de Physique
Th\'eorique, Universit\'e Libre de Bruxelles, CP225, Bd. du Triomphe, 1050
Bruxelles, Belgium. e-mail: fenglert@ulb.ac.be}, 
Laurent Houart
\footnote{Chercheur associ\'e CNRS and Charg\'e de Recherches FNRS \`a
titre honorifique, e-mail: lhouart@lpthe.jussieu.fr} 
and \
Paul~Windey
\footnote{e-mail: windey@lpthe.jussieu.fr}\\[.3cm]
{\it Laboratoire de Physique Th\'eorique et Hautes Energies
\footnote{ Laboratoire associ\'e No. 280 au CNRS.}\\ 
Universit\'e Pierre et Marie Curie, Paris VI\\ Bte 126,
4 Place Jussieu, 75252 Paris cedex 05, France}
\end{center}
\vskip 1cm
\begin{abstract}
\noindent The coupling of a Nambu-Goto string to gravity allows for
Schwarzschild black holes whose entropy to area relation is
$S=(A/4)(1-4\mu)$, where $\mu$ is the string tension.
\end{abstract}
\vskip1cm

It is well known that to compute thermal correlation functions and
partition functions in field theory in flat Minkowski spacetime one can
use path integrals in periodic imaginary time.  The period $\beta$ is the
inverse temperature and can be chosen freely.
  
This method was generalized to compute matter correlation functions in
static curved backgrounds.  For the Schwarzschild black hole, the analytic
continuation to imaginary time defines a Euclidean background everywhere
except at the analytic continuation of the horizon, namely the 2-sphere
at $r=2M$, $M$ being the black hole  mass. This ambiguity is usually removed by
assuming that the Euclidean manifold is regular at $r=2M$. The Euclidean
black hole  obtained in this way has a periodicity $\beta$ uniquely defined in
terms of the black hole  mass. The relation is
\begin{equation}
\label{mass}
\beta = 8\pi M.  
\end{equation} 
This value coincides with the temperature of the thermal quantum radiation
computed for the incipient black hole in the absence of back-reaction
\cite{haw1}.

In pioneering work, Gibbons and Hawking \cite{gh1} extended the analytic
continuation to the gravitational action, restricting the hitherto
ill-defined path integral over metrics to a saddle point in the Euclidean
section. To constitute such a saddle the Euclidean black hole must be
regular given that a singularity at $r=2M$ would invalidate the solution of
the Euclidean Einstein equations. The partition function evaluated on this
saddle is interpreted as $e^{-\beta F}$, where $F$ is the free energy of
the background black hole spacetime. It yields the Bekenstein-Hawking
\cite{bek} area entropy $S$ for the black hole namely
\begin{equation} 
\label{area}
 S = {A\over 4}   
\end{equation} 
where $A$ is the area of the event horizon. In what follows, we shall
assume the general validity of the Gibbons-Hawking Euclidean saddle
condition. Its significance will be further discussed elsewhere \cite{ehw}.

The relation Eq.\ (\ref{mass}) between the temperature and the black hole
mass is affected by classical surrounding matter but the entropy remains
unchanged and is still given by Eq.\ (\ref{area}). This value of the
entropy seems therefore to depend only on the black hole mass. In this
letter it will be shown that it does not. A different relation between
entropy and area will be presented when a conical singularity is present in
the Euclidean section at $r=2M$.

Many authors \cite{cpw,dgt,btz,tei,ct,su} have introduced a conical
singularity at $r=2M$. This modifies the Euclidean periodicity of the black
hole and therefore the temperature. However if the source producing this
singularity is not taken into account, the Euclidean black hole solution is
not a saddle point of the functional integral. A true Euclidean saddle
point can nonetheless be maintained by introducing an elementary string in
the action. The conical singularity arises from a string ``instanton'' and
the associated deficit angle is determined by the string tension.  The
temperature depends on the string tension and it now necessarily entails a
variation of the entropy versus area ratio. This ratio takes value in the
interval $[0,1/4]$, the lower limit being approached when the cone
degenerates. At fixed string tension, the relation between entropy and area
remains insensitive to the introduction of classical surrounding matter.

The Lorentzian action for gravity coupled to matter fields is taken to be
\begin{equation}
\label{L-action}
I={1\over 16\pi}\int_M\ \sqrt{\vert g\vert} R -{1\over 8\pi}\int_{\partial
M}\ \sqrt{\vert h\vert}K + I_{matter}.
\end{equation}
Here ${1\over 16\pi}\int_M\ \sqrt{\vert g\vert}R$ is the usual
Einstein-Hilbert action, $K$ is the trace of the extrinsic curvature on the
boundary $\partial M$ of the four dimensional manifold $M$, and $h$ the
determinant of the induced metric.

The introduction of the $K$-term requires explanation. We will justify it
briefly and refer the interested reader to the recent detailed discussion
of Hawking and Horowitz \cite{haho}. The Einstein-Hilbert action contains
second-order derivatives of the metric. If the system evolves between two
non intersecting spacelike hypersurfaces these second derivative terms can
be transformed by partial integration into boundary terms on these
spacelike surfaces and on timelike surfaces. Explicitly these boundary
terms stem from the integral of the four-divergence $\partial_{\mu}
{\omega}^\mu$ where
\begin{equation}
\label{omega}
\omega^\mu = -{1 \over 16\pi}\left( \partial_\nu( \sqrt{\vert g\vert} g^{\mu
\nu})+ g^{\mu \nu} \partial_\nu \sqrt{\vert g\vert}\right). 
\end{equation} 
Their contribution to the action Eq.\ (\ref {L-action}) is cancelled by the
$K$-term. The absence of boundary terms on the spacelike surfaces is
necessary for the consistency of the Hamiltonian formalism. However, for
the asymptotically flat spaces considered here, the $K$-term
introduces divergences at spacelike infinity.  These can be removed by
subtracting from Eq.\ (\ref {L-action}) a $K$-term at infinity in flat
space. It can then be verified that the subtracted action
\begin{eqnarray}
\nonumber
I-I_0&=&{1\over 16\pi}\int_M\ \sqrt{\vert g\vert} R -{1\over
8\pi}\int_{\partial M}\ \sqrt{\vert h\vert}K \\
\label{subtraction}
&&+ {1\over
8\pi}\int_{(\partial M)_\infty}\ \sqrt{\vert h_0\vert} K_0 + I_{matter}
\end{eqnarray}
yields the correct ADM mass as the on-shell value of the Hamiltonian. The
action Eq.\ (\ref {subtraction}) can now be written as a Hamiltonian action
and the path integral over metrics can be formally defined.
   
We will consider a system where the matter is an elementary Nambu-Goto
string. Its action is given by
\begin{equation}
\label{ng} 
I_{matter}\equiv I_{string}= -\mu \int\ d^2\sigma \sqrt{\vert\gamma\vert},
\end{equation} 
where $\mu$ is the string tension and $\gamma$ is the determinant of the
induced metric on the worldsheet: \begin{equation}
\gamma_{ab}(z)=g_{\mu\nu}(z)\partial_az^\mu\partial_b z^\nu.
\end{equation}
In the presence of a string, the Lorentzian Einstein equations still admit
ordinary black hole solutions corresponding to trivial zero string area.
The continuation of these solutions to imaginary time is given by the
metric
\begin{equation}
\label{schw}
ds^2=\left(1-{2M \over r}\right) dt^2+ \left(1-{2M \over r}\right)^{-1}
dr^2+r^2 d\Omega^2
\end{equation}
for $r > 2M$. To determine the Euclidean background at $r=2M$ we impose the
Euclidean Einstein equations. As previously discussed, the free energy will
then be computed on the saddle of the Euclidean action.
 
The action Eq.\ (\ref {L-action}) with matter term Eq.\ (\ref {ng}) can be
extended to Euclidean metrics.  The subtracted Euclidean action reads
\begin{eqnarray}
\nonumber
\widetilde I-\widetilde I_0&=&-{1\over 16\pi}\int_{\widetilde M}\ \sqrt{ g }
R +{1\over 8\pi}\int_{\partial \widetilde M}\ \sqrt{ h }K \\
\label{action}
&&- {1\over
8\pi}\int_{(\partial \widetilde M)_\infty}\ \sqrt{ h_0 } K_0 + \mu \int\
d^2\sigma \sqrt{\gamma}.
\end{eqnarray} 
Here the world sheet has the topology of a 2-sphere. Because the Euclidean
black holes Eq.\ (\ref{schw}) have only one boundary, namely at infinity,
we must take $\partial \widetilde M =(\partial \widetilde M)_\infty$ in the
$K$-term. The $K_0$-term subtraction has to be performed with the extrinsic
curvature in flat Euclidean space. The independent variables in Eq.\
(\ref{action}) are the components of the metric $g_{\mu\nu}$ and the string
coordinates $z^\mu$. The variation of the action with respect to
$g_{\mu\nu}$ gives the Euclidean Einstein equations:
\begin{equation}
\label{einstein}
R_{\mu\nu}(x)-{1 \over 2} g_{\mu\nu}(x)\ R(x)=8 \pi T_{\mu\nu} (x)
\end{equation} 
where 
\begin{equation}
T^{\mu\nu}(x)=-\mu\int d^2\sigma \sqrt{\gamma}\gamma^{ab}\partial_az^\mu
\partial_b z^\nu {1 \over \sqrt{ g(z)}} \delta^4 (x-z).
\end{equation} 
Variations with respect to $z^{\mu}$ give rise to the stationary area
condition for the string.

The Einstein equations Eq.\ (\ref {einstein}) still admit ordinary
Euclidean black hole solutions corresponding to zero string area.  The
Euclidean space is regular at $r=2M$ and the $t$-periodicity is
\begin{equation}
\label{period} 
\beta_H=8\pi M.
\end{equation}
However there exists a non-trivial solution to the string equations of
motion when the string wraps around the Euclidean continuation of the
horizon, a sphere at $r=2M$.  All solutions are correctly described by the
metric Eq.\ (\ref{schw}) but the non-trivial one has a curvature
singularity at $r=2M$. 

The trace of Einstein equations Eq.\ (\ref{einstein}) gives
\begin{equation}
\label{trace}
\int_{\widetilde M} \sqrt{  g } R = 16\pi \mu A,
\end{equation}
where $A$, the area of the string, is equal to the area of the horizon, a
two sphere at $r=2M$.  The entire contribution to the integral comes from
the singularity at $r=2M$. To evaluate $\int \sqrt{g} R$ when $R$ is zero
everywhere except at $r=2M$, one can consider an infinitesimal tubular 
neighborhood $S^2
\times D$ of $r=2M$ \cite{btz,tei}. This gives
\begin{equation}
\label{r4asr2}
\int_{S^2 \times D} \sqrt{\ g}\ R = A\int_{D}
\sqrt{ ^{(2)}g}\ ^{(2)}\!R.
\end{equation}
From Eq.\ (\ref{trace}) we have 
\begin{equation}
{1\over 4\pi}\int_{D}\  \sqrt{ ^{(2)}g} \, ^{(2)}R = 4\mu.
\end{equation}
This result and the Gauss-Bonnet theorem for disc topology tell us that
there is a conical singularity with deficit angle $ 2 \pi \eta$ such that
\begin{equation}
\label{deficit}
\eta=4 \mu.
\end{equation}
This deficit angle is the sole effect of the string instanton. 
The periodicity in $t$ is now:
\begin{equation}
\label{temp}
\beta=\beta_H \ (1-4 \mu).
\end{equation}
The string instanton has raised the global temperature from $\beta^{-1}_H$
to $\beta^{-1}$.
 
We now evaluate the free energy of the black hole. The contribution of the
string term to the action Eq.\ (\ref{action}) exactly cancels the
contribution of the Einstein term as seen from Eq.\ (\ref{trace}). The
boundary terms at asymptotically large $r=r_{\infty}$ are thus the only
ones contributing to $\beta F$. Using Eqs.\ (\ref{omega}) and (\ref{schw})
we find
\begin{equation}
\label{kterm}
{1\over 8\pi}\int_{\partial \widetilde M=(\partial \widetilde M)_\infty}\ \sqrt{h}K=-\beta 
\left(r_{\infty}\left(1-{2M \over r_{\infty}}\right) +{M
\over 2}\right).  
\end{equation}
The subtracted term is computed similarly in the flat metric
\begin{equation}
\label{flat}
ds^2 = \left(1-{2M \over r_{\infty}}\right) dt^2 + dr^2+ r^2 d\Omega^2,
\end{equation}
where $t$ has the periodicity $\beta$ given by Eq.\ (\ref{temp}).
The subtraction term is 
\begin{equation}
\label{kflat}
{1\over 8\pi}\int_{(\partial \widetilde M)_\infty}\ \sqrt{h_0}K_0=-\beta
r_{\infty} \left(1-{2M \over r_\infty}\right)^{1 \over 2}.
\end{equation}
The free energy is given by 
\begin{equation}
\label{free}
F = \beta^{-1} (\widetilde I-\widetilde I_0)_{saddle} = {M \over 2}.
\end{equation}
Note that the free energy has the same value it had in the absence of the
string instanton. However, using Eq.\ (\ref{temp}), the entropy
\begin{equation}
\label{defe} 
S = \beta^2 {dF \over d\beta}={\beta^2
\over 2}
{dM \over
d\beta}
\end{equation}
is now given by
\begin{equation}
\label{entropy}
S=(1-4 \mu) {A \over 4}.
\end{equation}
This is our central result. The introduction of the string has enabled us
to define a black hole with fixed mass $M$ at a temperature other than the
usual $\beta_H^{-1}$. When the temperature is not the Hawking temperature,
the entropy changes from $A/4$ to Eq.\ (\ref{entropy}).

It follows from Eq.\ (\ref{temp}) and Eq.\ (\ref{entropy}) that the product
of the entropy and the temperature is constant for a given mass,
independent of the string tension.
\begin{equation}
\label{state}
\beta^{-1} S= \beta^{-1}_H S_H = {M\over 2},
\end{equation}
where $S_H=A/4$.

To complete the thermodynamic analysis, we verify using
Eqs.(\ref{free}) and (\ref{state}) that the energy of
the solution is unchanged by the presence of the string instanton.
\begin{equation}
\label{energy}
E= F + \beta^{-1} S=M.
\end{equation}

It can be shown \cite{ehw} that the entropy Eq.\ (\ref{entropy}) is not
affected by the presence of classical matter surrounding the black hole.
In this case, Eq.\ (\ref{state}) becomes
\begin{equation}
\label{gestate}
\beta^{-1} S =\beta_{Hm}^{-1} S_H,
\end{equation}
where $\beta_{Hm}^{-1}$ is the inverse Hawking temperature in presence of
matter. This indicates that the entropy is a genuine property of the
horizon.

We now comment about the interpretation of our result from the Lorentz\-ian
viewpoint. As stated in the introductory paragraphs of this Letter, there
is an ambiguity in making the analytic continuation to imaginary time at
the horizon of the Lorentzian black hole metric. This ambiguity was removed
by requiring that the Euclidean metric be a saddle point of the Euclidean
action. This requirement imposes that one considers all available
saddles. The solution with the string instanton is one such
saddle\footnote{Multi-instantons could be considered. They have different
temperature and correspond to distinct thermodynamic states. Note however
that the deficit angle cannot exceeds its maximal value of $2\pi$ for which
the cone degenerates.} . It corresponds to the original Lorentzian black hole in
a different thermodynamic state than the solution with no wrapping. This
new state is characterized by a different temperature than the Hawking
temperature and therefore by different boundary conditions for quantum
fields in the Lorentzian background. It is well known that the Hawking
temperature corresponds to the regularity of the expectation value of the
energy momentum tensor of quantum fields in the Lorentzian background
\cite{cand}. Therefore the new boundary conditions, corresponding to the
temperature determined by the string instanton, lead to well defined
singularities in the energy momentum tensor of quantum fields on the
horizon. This will be discussed in more detail in a separate publication
\cite{ehw}. These singularities are quantum effects in a usual Schwarzschild
background which is regular on the horizon. Of course, these considerations
do not take into account the backreaction.

In this Letter, we have shown that the area entropy of an eternal
Schwarz\-schild black hole in the presence of a string instanton differs from
the usual value $A/4$ and depends on the value of the string tension.  Such
a black hole differs from an ordinary Schwarzschild black holes only by the
instanton effect. One cannot distinguish between them through their mass or
even through their Lorentzian metrics. They differ through quantum effects,
not in classical quantities such as $M/2$, the product of $\beta^{-1}$ and
$S$. Indeed the temperature is proportional to $\hbar$ while the entropy is
inversely proportional to $\hbar$. In this sense the instanton provides a
quantum hair \cite{cpw} which affects the expectation values of operators.
The string selects from all possible black holes of mass $M$ a subset
distinguishable by quantum effects and characterized by a smaller
entropy. This fact give credence to the interpretation of the area entropy
as a counting of states and points towards a possible retrieval by quantum
effects of the information concealed by, or stored in, the black hole
horizon.

We would like to thank R.~Parentani for very interesting comments on black
hole thermodynamics during a Workshop held at Paris VI.  We are grateful to
the other participants, R.~Argurio, L.~Baulieu, C.~Bouchiat, M.~Henneaux,
J.~Iliopoulos, S.~Massar, M.~Picco, and Ph.~Spindel, for valuable
discussions.  One of us (P.W.) would like to thank M.~O'Loughlin and
E.~Martinec for many interesting conversations.  This work was supported in
part by the Centre National de la Recherche Scientifique and the EC Science
grant ERB 4050PL920982.


\begin{thebibliography}{99}

\bibitem{haw1} S.W.~Hawking, Commun. Math. Phys. {\bf 43} (1975) 199.
\bibitem{gh1}G.W.~Gibbons and S.W.~Hawking,  Phys. Rev. D {\bf 15} (1977) 2752.
\bibitem{bek} J.D.~Bekenstein,  Phys. Rev. D  {\bf 7} (1973) 2333.
\bibitem{ehw} F.~Englert, L.~Houart and P.~Windey,
``Black Hole Entropy and String Instantons,'' 
Report No PAR-LPTHE 95-38 and ULB-TH 10/95, hep-th/9507061, 1995. 
\bibitem{cpw} S.~Coleman, J.~Preskill and  F.~Wilczek, Nucl. Phys.
 {\bf B378} (1992) 175.
\bibitem{dgt} F.~Dowker, R.~Gregory and J.~Traschen, Phys. Rev. D {\bf 45}
 (1992) 2762.
\bibitem{btz} M.~Ba\~nados, C.~Teitelboim and J.~Zanelli,
 Phys. Rev. Lett. {\bf 72} (1994) 957.
\bibitem{tei} C.~Teitelboim, Phys. Rev. D  {\bf 51} (1995) 4315.  
\bibitem{ct}  S.~Carlip and C.~Teitelboim, Class. Quant. Grav.
{\bf 12} (1995) 1699. 
\bibitem{su}  L.~Susskind and J.~Uglum, Phys. Rev. D {\bf 50}
(1994) 2700.
\bibitem{haho} S.W.~Hawking and G.~Horowitz,  
``The Gravitational Hamiltonian, Action, Entropy and Surface Terms,''
Report No DAMTP/R 94-52 and UCSBTH-94-37, gr-qc/9501014, 1995 .
\bibitem{cand} P.~Candelas, Phys. Rev. D {\bf  21} (1980) 2185.
\end{thebibliography}
\end{document}